\documentclass{emulateapj}

\begin{document}

\newcommand{\fac}{2}
\newcommand{\ntot}{24}
\newcommand{\nd}{18}
\newcommand{\nspec}{15}
\newcommand{\fbar}{$2.3 \pm 3.6 \times10^{-17}\ {\rm erg\ s^{-1}\  cm^{-2}}$}
\newcommand{\sfrbar}{$24 \pm 6\ {\rm M_{\odot}\  yr^{-1}}$}
\newcommand{\fbarhb}{$4.4 \pm 3.1 \times10^{-18}\ {\rm erg\ s^{-1}\ cm^{-2}}$}
\newcommand{\frat}{$5.1^{+0.5}_{-0.5} $}
\newcommand{\nall}{49}
\newcommand{\ncomb}{67}
\newcommand{\fratall}{$4.8^{+0.8}_{-1.7}$}
\newcommand{\g}{$g_{475}$}
\newcommand{\B}{$B_{435}$}
\newcommand{\V}{$V_{606}$}
\newcommand{\ra}{$r_{625}$}
\newcommand{\ia}{$i_{775}$}
\newcommand{\I}{$I_{814}$}
\newcommand{\Hw}{$H_{160}$}
\newcommand{\z}{$z_{850}$}
\newcommand{\K}{$K$}
\newcommand{\msun}{${\rm M_{\odot}}$}
\newcommand{\sfrun}{${\rm M_{\odot}\ yr^{-1}}$}

\newcommand{\etal}{{\em et~al.\,}}

\title{Rest-Frame Optical Emission Lines in $z\sim3.5$ Lyman Break selected
  Galaxies: The Ubiquity of Unusually High [OIII]/H$\beta$ Ratios at 2 Gyr
  \altaffilmark{1,2}}

\altaffiltext{1}{The data presented herein were obtained at
  the W.M. Keck Observatory, which is operated as a scientific
  partnership among the California Institute of Technology, the
  University of California and the National Aeronautics and Space
  Administration. The Observatory was made possible by the generous
  financial support of the W.M. Keck Foundation.}

\altaffiltext{2}{Partially based on data obtained with the
  \textit{Hubble Space Telescope} operated by AURA, Inc. for NASA
  under contract NAS5-26555. Partially based on observations  with the
  \textit{Spitzer Space Telescope}, which is operated by the Jet
  Propulsion Laboratory, California Institute of  Technology under
  NASA contract 1407. }

\author{B. P. Holden\altaffilmark{3}}
\author{P. A. Oesch\altaffilmark{3,4}}
\author{V. G. Gonz{\'a}lez \altaffilmark{5}}
\author{G. D. Illingworth \altaffilmark{3}}
\author{I. Labb{\'e} \altaffilmark{6} }
\author{R. Bouwens \altaffilmark{6}}
\author{M. Franx\altaffilmark{6} }
\author{P. van Dokkum\altaffilmark{4} }
\author{L. Spitler \altaffilmark{7,8} }

\altaffiltext{3}{UCO/Lick Observatories, University of California,
  Santa Cruz, 95065; holden@ucolick.org, gdi@ucolick.org}
\altaffiltext{4}{Yale Center for Astronomy and Astrophysics, Dept. of Astronomy, Yale University; pascal.oesch@yale.edu, pieter.vandokkum@yale.edu}
 \altaffiltext{5}{Dept. of Astronomy, University of California,
   Riverside; valentino.gonzalez@ucr.edu }
\altaffiltext{6}{Leiden Observatory, Leiden University,  P.O. Box 9513, 2300 RA Leiden,The Netherlands; ivo@strw.leidenuniv.nl,  franx@strw.leidenuniv.nl}
\altaffiltext{7}{Dept. of Physics and Astronomy, Macquarie University; lee.spitler@mq.edu.au}
\altaffiltext{8}{Australian Astronomical Observatory, North Ryde, Australia}

\begin{abstract}

  We present \K-band spectra of rest-frame optical emission lines for
  \ntot\ star-forming galaxies at $z\sim3.2\mbox{--}3.7$ using MOSFIRE
  on the Keck 1 telescope.  Strong rest-frame optical [\ion{O}{3}] and
  H$\beta$ emission lines were detected in \nd\ LBGs. The median flux
  ratio of [\ion{O}{3}]$\lambda$5007 to H$\beta$ is \frat.  This is a
  factor of $5\mbox{--}10\times$ higher than in local galaxies with
  similar stellar masses. None of our sources are detected in deep
  X-ray stacks, ruling out significant contamination by active
  galactic nuclei. Combining our sample with a variety of LBGs from
  the literature, including \nall\ galaxies selected in a very similar
  manner, we find a high median ratio of [\ion{O}{3}]/H$\beta = $
  \fratall. This high ratio seems to be an ubiquitous feature of
  $z\sim3\mbox{--}4$ LBGs, very different from typical local
  star-forming galaxies at similar stellar masses.  The only
  comparable systems at $z\sim0$ are those with similarly high
  specific star-formation rates, though $\sim5\times$ lower stellar
  masses. High specific star-formation rates may result in a higher
  ionization parameter, higher electron density, or harder ionizing
  radiation, which, combined different elemental abundances, result
  in a much higher [\ion{O}{3}]/H$\beta$ line ratio. This implies a
  strong relation between a global property of a galaxy, the specific
  star-formation rate, and the local conditions of ISM in star-forming
  regions.
\end{abstract}

\keywords{galaxies: evolution  --- galaxies: high-redshift --- galaxies: spectroscopy  }

\section{Introduction}

Tracing out the star-formation history of the universe is a key
ingredient for our understanding of the mass assembly of
galaxies. Great progress has been made in the last decade or so based
on deep imaging surveys both from the ground and from space with the
Hubble Space Telescope (\emph{HST}). These have led to the very
efficient identification of several thousand star-forming galaxies at
$z\geq4$ based on broad-band imaging, using the Lyman break selection
technique \citep[e.g., ][]{steidel1996}. Thanks to the combination of
\emph{HST} and \emph{Spitzer}/IRAC imaging, the analysis of these
galaxies was further extended from rest-frame UV only studies to
include the rest-frame optical, leading to reliable estimates of the
stellar mass functions of very faint galaxies out to $z\sim7-8$
\citep[e.g., ][]{stark2009,Labbe10b, gonzalez2012, lee2012}.

Since some of the strongest emission lines such as H$\alpha$ or
[\ion{O}{3}]$\lambda$5007 are shifted into the observed-frame
near-infrared at $z\gtrsim2$, progress on spectroscopic confirmation and
observation of rest-frame optical lines was very slow and time
consuming. However, with the advent of efficient multi-object
near-infrared spectrographs on 8m-class telescopes, this situation is
now changing. In this paper we present an analysis of [\ion{O}{3}]
and H$\beta$ emission lines of a sample of $z\sim3.5$ galaxies based on
observations with the Keck MOSFIRE instrument \citep{mcclean2012}.

The strength of such lines is very important as they can provide key
insight into the conditions of star-formation in high-redshift
sources, and when combined with additional line measurements such as
[\ion{O}{2}] can provide an estimate of the gas-phase metallicity of
$z\sim3-4$ galaxies
\citep[e.g.][]{maiolino2008,mannucci2009,troncoso2013}.

There is now growing evidence for a high fraction of galaxies showing
strong nebular line emission at $z\sim4-8$, with median rest-frame
equivalent widths (EW$_0$) of $\gtrsim300$ \AA \citep[e.g.,
][]{schaerer2009,debarros2012,stark2013}.  From a sample of 74
isolated galaxies with deep $Spitzer$/IRAC photometry and with
spectroscopic redshifts in the range $3.8<z<5.0$, \citet{shim2011}
find that 65\% show clear flux excess in IRAC [3.6], indicative of
strong H$\alpha$ emission. From this excess in the broad-band
photometry, they estimate rest-frame equivalent widths in the range
$140 - 1700$ \AA. In a similar analysis, \citet{stark2013} derive the
EW(H$\alpha$) distribution at this redshift, finding a mean value of
270 \AA. \citet{gonzalez2012} find a similar flux excess in the median
stacked SEDs of galaxies at $z\sim5,~6,$ and 7 \citep[see
also][]{labbe2012, smit13}. This excess suggests that [OIII] and
H$\beta$ also have large rest-frame EWs and that strong emission lines
may be ubiquitous over a range of masses ($M=10^9\mbox{--}10^{10}$
M$_\odot$). Most recently, \citet{schenker2013} used a sample of 20
Lyman Break galaxies (LBGs) to spectroscopically confirm that the
majority of $z\sim3.5$ galaxies have strong [\ion{O}{3}] equivalent
widths in agreement with the H$\alpha$ equivalent width distribution
based on the broad-band IRAC photometry.

In the local universe, galaxies with extreme emission lines have been
identified from SDSS based on extremely blue $r-i$ colors because the
$r$-band is dominated by the [OIII] line
\citep[e.g.]{kakazu2007,cardamone2009}. Such low-metallicity
starbursts only contribute a small fraction of the total
star-formation at $z<1$. At higher redshift, $z\sim1.7$, similar
observations based on extreme broad band colors have revealed a
significant population of galaxies that are undergoing vigorous
star-formation episodes. This is indicated by their large [OIII] +
H$\beta$ EWs $\sim 1000\,$\AA, that imply that they can build their
whole stellar mass in only 15 Myr
\citep{vanderwel2011,maseda2013}. Similarly strong line emitters have
independently been found even up to $z\sim2.3$ from WFC3/IR grism
spectroscopy \citep{atek2011,xia2012}.  At these higher redshifts, the
number density of such sources is found to be quite significant
($\sim4\times10^{-4}$ Mpc$^{-3}$).

These strong emission lines point to unusual properties in the high
redshift star-forming population. As an example, one can consider the
Lynx arc, which in many ways is a prototypical galaxy at high redshift
with unusual properties.  \citet{fosbury2003} studied this object with
deep, high quality spectra across the rest-frame UV and optical. The
galaxy shows strong ionization, such that the [\ion{O}{2}] line is not
detected despite the detection of the [\ion{Ne}{3}] at 3869 \AA. To
reproduce these line strengths and ratios, a cluster of $\sim10^{5}$
stars with surface temperatures of 80,000 K is required. An absorbed
AGN could explain this source, though that possibility makes specific
predictions \citep{binette2003}.

It appears, however, that such objects are common at z$\sim$2
\citep{nakajima2013} requiring different interstellar medium
properties than we see in the local universe \citep{kewley2013a},
potentially with significantly higher ionized gas densities
\citep[e.g.][]{shirazi2013}. Recent theoretical models can reproduce
such line ratios, but they require very different ionization
parameters, strong winds, radiation pressure or significant shocks
even at the low metallicity values expected for galaxies at such an
early epoch in the universe \citep{yeh2013,verdolini2013,kewley2013b,rich2014}.

In this paper, we probe the star-formation properties of a sample of
$z\sim3.5$ star-forming galaxies with the use of Keck MOSFIRE
multi-object near-infrared spectroscopy to target the [\ion{O}{3}] and
H$\beta$ lines with a single mask observation in the GOODS-South
field.  The paper is organized as follows: in Section
\ref{sec:selection}, we present our target selection, before
describing our observations in Section \ref{sec:obs}, and outlining
our analysis in Section \ref{sec:analysis}. Finally, we end with the
presentation and a discussion of our results in Section
\ref{sec:results}.  Throughout this paper, we adopt AB magnitudes and
a standard cosmology with $\Omega_M=0.3, \Omega_\Lambda=0.7$, and
$H_0=70$ km~s$^{-1}$Mpc$^{-1}$.

\section{Target Selection and Photometric Data}
\label{sec:selection}

Our primary goal is to investigate the rest-frame optical emission
line properties of Lyman break selected, star-forming galaxies at
$z\sim3.5$. The GOODS-South field offers the best combination of
multi-wavelength imaging data both from the ground and from space with
HST and Spitzer, as well as a large sample of spectroscopically
confirmed LBGs in the required redshift range based on the large
campaign of rest-frame UV spectra
\citep[e.g.][]{vanzella2008,balestra2010}.

\subsection{Pre-Existing Spectroscopic Sample}
\label{presamp}

The primary emission lines of our study are H$\beta$ and the
[\ion{O}{3}]$\lambda\lambda$4959,5007 doublet.  These lines are
accessible in the \K\ band over the redshift range $3.2 < z < 3.8$.  To
increase the efficiency of our observations, we prioritized galaxies
with existing spectroscopic redshifts such that their H$\beta$ and
[\ion{O}{3}] lines would fall in between the many strong night sky
lines. To do this, we used the spectroscopic catalogs of GOODS-South
compiled by \citet{vanzella2008} and \citet{balestra2010} who observed
large samples of LBGs in this redshift range. 

We identified an overdensity of galaxies in the South-West part of
GOODS-South with existing spectroscopic redshifts for which our target
lines would lie in between sky lines, and we thus chose to center our
mask design around that apparent overdensity. We designed the mask to
maximize the number of galaxies with known redshifts and were able to
fit \nspec\ into our design. 

The rest-frame UV spectra of these galaxies show Lyman $\alpha$ in
emission in 8 of the \nspec\ galaxies, with measured equivalent widths
ranging from 4 to 72 \AA\ \citep[measured form the spectra of
][]{balestra2010, vanzella2008}.

In addition to the spectroscopically confirmed sample, we included nine
\B\ dropout galaxies without pre-existing rest-frame UV redshift
measurements as secondary targets. 

Since our input spectroscopic redshift samples are based on rest-frame
UV spectra, we are selecting galaxies bright in the rest-frame UV,
typically brighter than $i_{775}<25$ mag AB \citep{balestra2010}. These
magnitudes imply star-formation rates of $\gtrsim$10 \sfrun\ which
should yield solid detections for H$\beta$ (signal to noise $>$7) in a 3
to 4 hour exposure with the MOSFIRE spectrograph. The filler galaxies
were in general much fainter than the spectroscopic input sample.  The
final mask design included \ntot\ high-redshift targets, in addition to
one star for measuring the Telluric absorption.

\subsection{Imaging Data and Photometry}

\begin{figure*} 
\begin{center}\includegraphics[width=.8\linewidth]{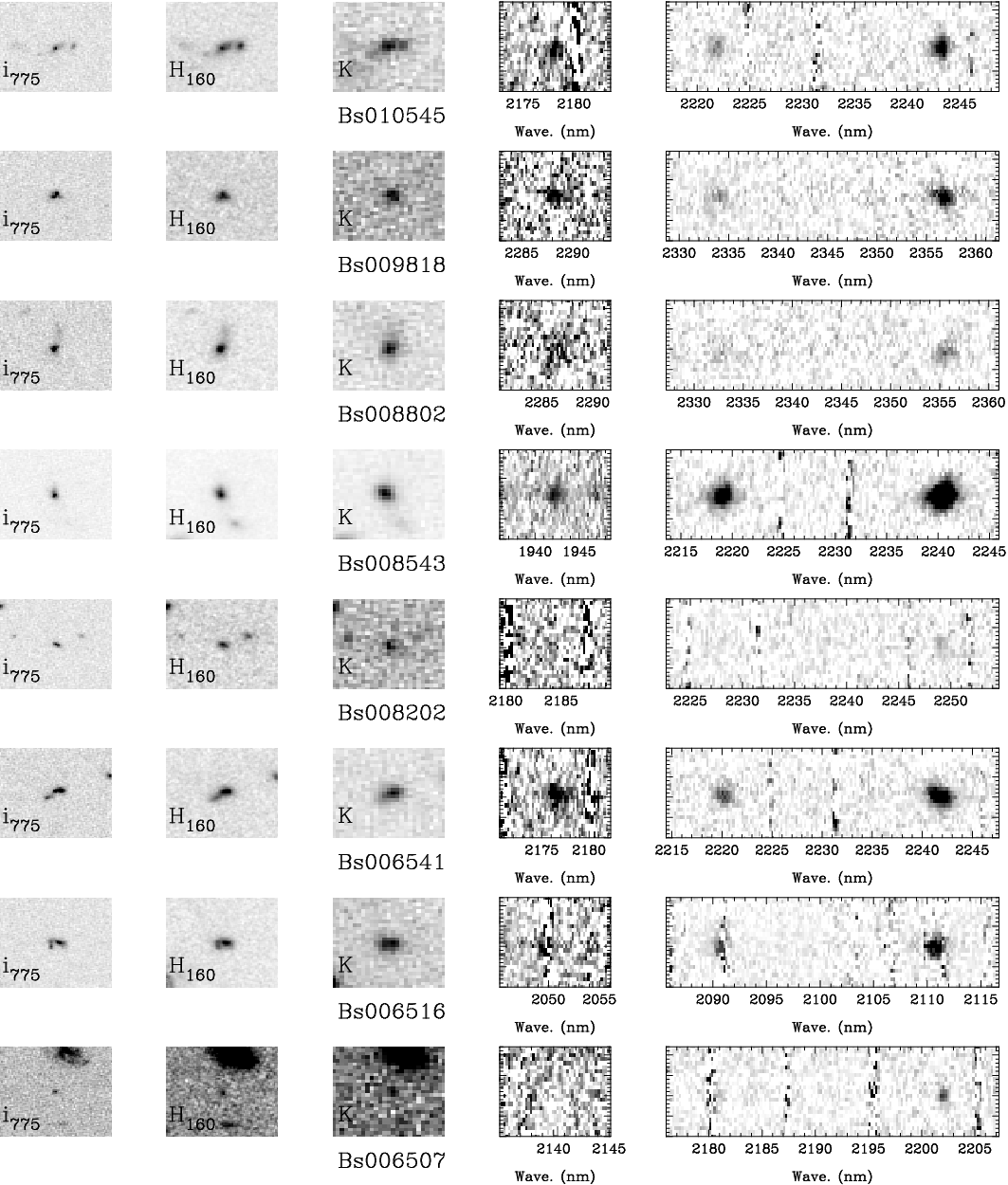} \end{center}
\caption{Images and
    MOSFIRE spectra of eight of the $z\sim3.2-3.8$ primary target galaxies.
    Each image is 4\arcsec on a side and was rotated such that the slit runs
    straight up along the y-axis. The spectra cover the two regions where H$\beta$
    and the [\ion{O}{3}] doublet can be found.}
\label{Bstamps}

\end{figure*}

\begin{figure*} 
\begin{center}\includegraphics[width=.8\linewidth]{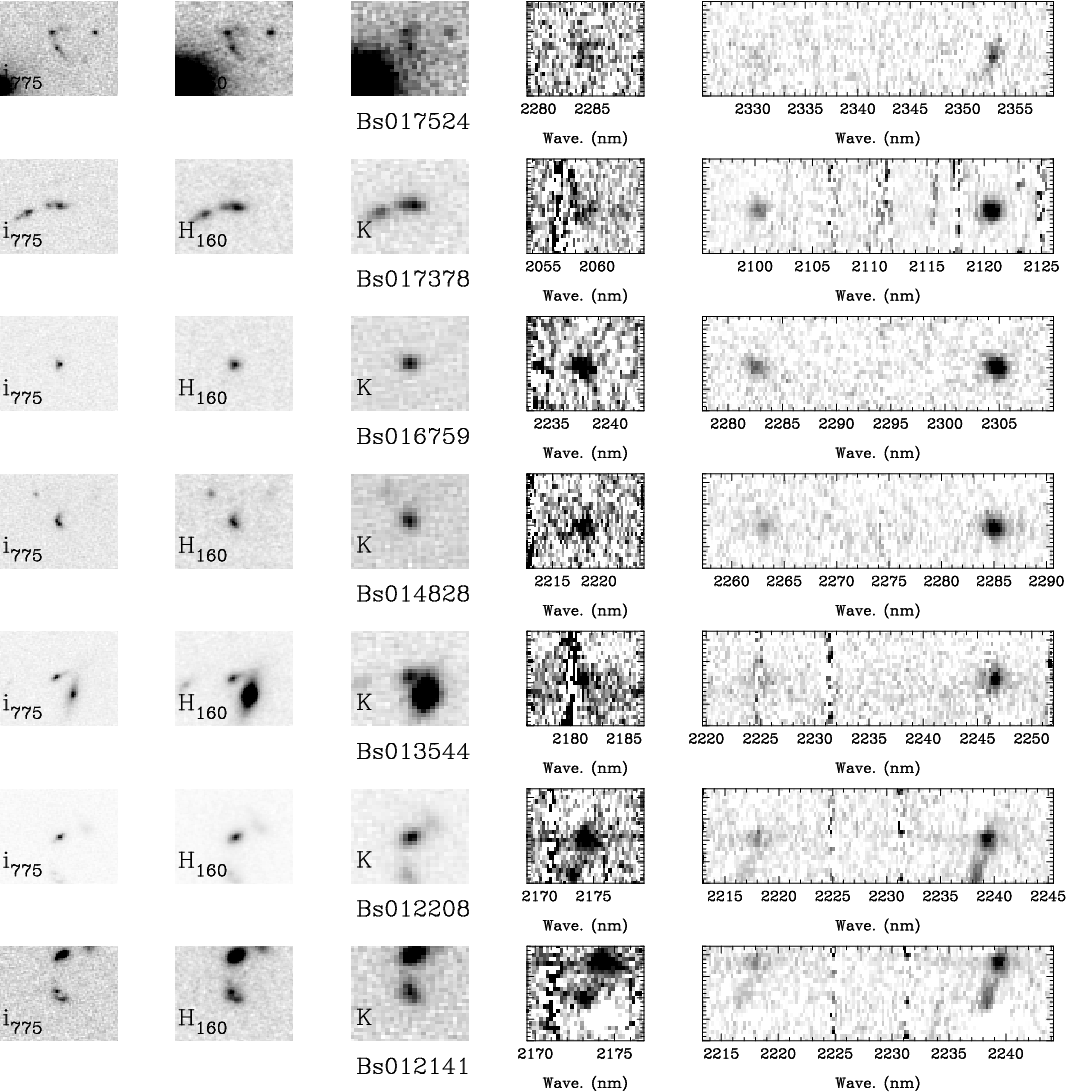} \end{center}
\caption{As in
    previous Figure, Images and MOSFIRE spectra of the remaining seven primary
    targets.}
\label{Bstamps2}

\end{figure*}

\begin{figure*}[htb]
\begin{center}\includegraphics[width=.8\linewidth]{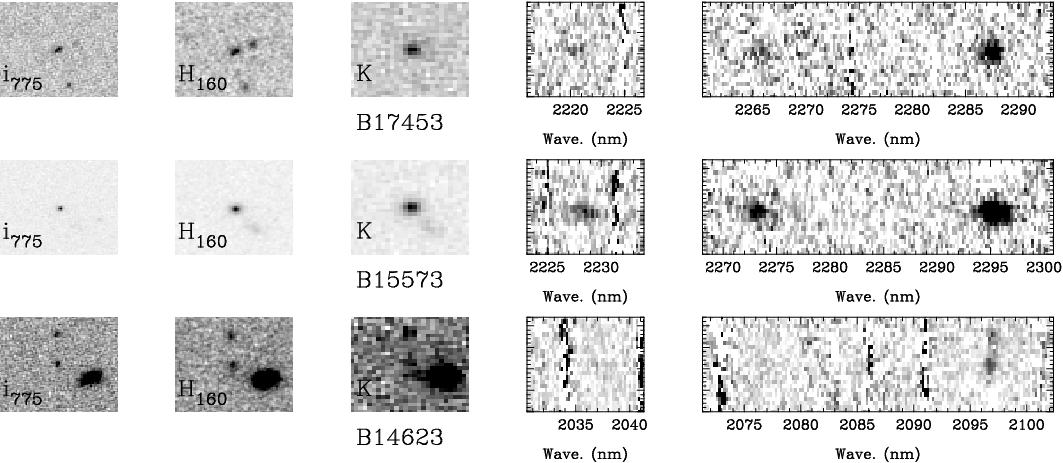}\end{center}
\caption{As in Figure \ref{Bstamps}, images and MOSFIRE spectra of the three \B-dropout galaxies,
  without pre-existing spectra, where we detected emission lines.}
\label{Bfillers}

\end{figure*}

All our targets are covered by multi-wavelength imaging data from
HST. We use our own reduction of the GOODS-South ACS data, which includes
additional follow-up observations and is therefore somewhat deeper than the
publicly available v2.0 images \citep{Giavalisco04b}, reaching to $i_{775} =
28.2$ mag AB (5$\sigma$ measured in small circular apertures of 0\farcs25
diameter). Additionally, we reduced all the WFC3/IR data from the complete
CANDELS GOODS-South imaging program \citep[PI: Faber/Ferguson;][]{grogin2011,
  koekemoer2011}, reaching a varying depth of $H_{160} = 27.5 - 28.3$ mag AB.
All our sources are $>1$ mag brighter than these limits, and therefore seen at
high significance in these images (see Figures \ref{Bstamps}, \ref{Bstamps2},
and \ref{Bfillers}). HST photometry is measured on PSF-matched images in small
Kron apertures, and is corrected to total fluxes using the $H_{160}$ band
image.

Imaging in the \K\ band was particularly important for our analysis, since this
is the band in which we took spectroscopic observations. We used a very deep
stack of \K\ imaging data consisting of a combination of all available data
over the CDF-S. This includes ESO/VLT ISAAC and HAWK-I data, along with PANIC
data from Magellan. The final image has exquisite seeing of only 0\farcs4. The
total \K-band photometry was measured in 1\farcs5 diameter apertures and was
corrected to total fluxes using the observed profiles of stars in the image. We
find a limiting magnitude of 26.2 mag AB for a 5$\sigma$ detection within a 0\farcs4
diameter aperture. The actual aperture photometry was performed using the sinc
interpolation procedure from \citet{bickerton2013}. To estimate the errors on
our fluxes, we placed 1000 random apertures scattered throughout the region of
the image our targets occupy.

For reliable mass estimates, we additionally measured rest-frame optical
photometry at longer wavelengths in the deep $Spitzer$/IRAC \citep{Fazio04}
[3.6] and [4.5] imaging data over GOODS-South \citep[][]{Dickinson03}.
Due to the large IRAC point-spread function (PSF), we used a sophisticated
neighbor subtraction scheme based on a convolution of the $J_{125}$ images to
the IRAC PSF. We then perform aperture photometry on the cleaned images in
2\arcsec\ diameter apertures, and correct to total fluxes using the growth
curves of nearby stars in the field \citep[for more information on the IRAC
photometry see][]{Labbe10b,Labbe10a,Oesch13,labbe2015}.

In Figures \ref{Bstamps}, \ref{Bstamps2} and \ref{Bfillers}, we show the
\ia, \Hw\ and \K\ data for each galaxy in our sample with a detected
emission line.

\section{Spectroscopic Observations}
\label{sec:obs}

We observed on January 4, 2013 for a single half night in good
conditions with the MOSFIRE instrument. We targeted a single mask in
the \K-band while the CDFS was visible from the Keck I telescope,
yielding a total exposure of 204 minutes with an AB dither pattern.
For the offsets between exposures we used both a 1\farcs 0 and 1\farcs
2 dither, in order to minimize the impact of potential bad pixels. The
seeing in the final stacked spectrum was 0\farcs 7.

\subsection{Reduction and Extraction}

The two-dimensional data reduction was performed using a slightly modified
version of the MOSFIRE DRP\footnote{https://code.google.com/p/mosfire/}. This
pipeline yields two-dimensional, sky-subtracted data for each slitlet, which
are rectified and wavelength calibrated, in units of electrons per second.
Because MOSFIRE has such high quality optics, the pixel size is almost constant
across the field of view and there is little other distortion, so the final
rectification is minor. The reductions for the two dither patterns, 1\farcs 0
and 1\farcs 2, were done separately. The two reductions were then averaged,
weighted by the exposure time for each separate stack.

All the 2D frames were searched for emission lines by eye, and their fluxes
were measured based on an optimal extraction \citep{horne1986} using a
Gauss-Hermite model. The line model was derived from the brightest line of a
given galaxy (i.e. [\ion{O}{3}]$\lambda$5007) and was then used to extract the
flux of the remaining lines. This was particularly important for lines that
were either very faint or sat partially on a night sky line.

In the two right-hand columns of Figures \ref{Bstamps}, \ref{Bstamps2}
and \ref{Bfillers}, we show the two-dimensional sky-subtracted spectra
for each galaxy with a detected emission line. The sky-subtracted
spectra are only shown in the wavelength regions of H$\beta$ and the
[\ion{O}{3}] doublet.

\subsection{Flux Calibration}
\label{cal}

For the overall zero-point of the spectra, we observed GD71 as spectrophotometric
standard. This observation was done with GD71 at an airmass of 1.04 for 120
seconds. This allows us to measure the conversion between $e^-{\rm s^{-1}}$ and
${\rm erg\ s^{-1}\ cm^{-2} \AA^{-1} }$, for that exposure. 

The above procedure does not deal with the variable seeing and airmass for the
observations of our mask. We correct this with a simple average flux correction
based on a single star within our mask. The magnitude of the star as computed
from the flux calibrated spectrum is $K = 18.24$ mag AB, while its total
(5\arcsec\ aperture) magnitude is $K = 18.09$ mag AB. This difference of 0.15
mag represents our estimated amount of flux lost between the typical
observation of the CDFS in a 0\farcs7 slit and the ideal observation of a star
at almost zenith. Therefore, we multiply all of our count rate fluxes by 1.25.

As we have only one flux calibration measurement, we do not have a good
estimate of the uncertainties on our calibration. \citet{schenker2013} found an
uncertainty of 15\% using a similar approach as we use. Unlike that work, we
will not include that uncertainty in our measurements errors.

We tabulate our results in Table \ref{spectab}. This table includes
the positions of the objects, the line fluxes without the aperture
correction, and the redshift of the object based on the flux weighted
centroids of the emission lines. The errors for the fluxes will be
discussed in the next section.

\subsection{Error Estimates}
\label{errors}

We estimated the errors for our spectra by simulations. We perform two
sets of simulations, one to estimate the statistical errors and a
second to estimate the error on the aperture correction. We simulate the flux
measurement procedure (\S \ref{fluxsim}) and use simulations to estimate
our aperture correction (\S \ref{apersim}). As a final check, we use the MOSFIRE
exposure time calculator to estimate the theoretical maximum of our
signal to noise (\S \ref{etcsim}).

The simulations rely on using the \K\ imaging data to make
templates of our spectroscopic observations. These templates are
smoothed to match the seeing in our spectroscopic data (0\farcs 7) as
compared to the 0\farcs4 seeing of the \K\ imaging. These were used to
generate artificial images of the lines. These artificial images were further
smoothed in the wavelength direction to match the sizes of the lines
in the data. 

\subsubsection{Simulations of the Flux Measurement}
\label{fluxsim}

Our first set of simulations use the simulated images to compute the
error from the model fitting. We place these simulated
images in actual data. We placed the simulated emission line at the
same wavelength but in other spectra. 

The simulated object is normalized to have the same flux as the object
was detected with in the spectrum. We then follow the same procedure
to estimate the flux, fitting the same order Gauss-Hermite polynomials
to model the 5007\AA\ line, and then use that same model to extract
the 4959\AA\ and H$\beta$ lines.
 
In general, we find for each object that the modeling process and
background variation errors are larger than accounted for in our
statistical errors. How much larger, however, varies depending on how
close the emissions are to the night sky background. For example, the
target B15573 has statistical errors of 2\%, 6\% and 7\% for 5007\AA,
4959\AA, and H$\beta$, respectively. The simulations show 3\%, 6\% and
10\% for the same lines. The larger error for H$\beta$ comes about
from the night sky lines near the emission line, as can be seen in
Figure \ref{Bfillers}. We tabulate these errors as the
statistical errors in Table \ref{spectab}.

\citet{kriek2015} presents  measurements of  the signal-to-noise  as a
function of flux measured for MOSDEF, a survey with MOSFIRE of a large
sample of galaxies, including a subset at the same redshift range
our sample. In general, the signal-to-noise of the sample from
\citet{kriek2015} is lower, as expected given the lower typical
exposure time for MOSDEF, a 120 minutes compared with our 204
minutes. In general, our results are $\sim$50\% in slightly less than
twice the exposure time, which is roughly consistent, for the emission
lines of interest, with a number of lines in the  \citet{kriek2015}
having signal-to-noise values higher than our data with less exposure
time. From this we conclude that our data are somewhat better than
average, but well within the locus shown in Figure 9 of \citet{kriek2015}.

\subsubsection{Simulations of the Aperture Correction}
\label{apersim}

Previous work has found that aperture corrections can be large, even
up to a factor of $\sim$2 \citep{erb2006c}. We use similar simulations as
above to estimate our uncertainties in our aperture corrections. We
take the deep \K\ imaging and smooth it to match the seeing in our
MOSFIRE spectra. We then place apertures on object at multiple
angles. We sum the flux from the object in the aperture and compare
that to the total flux. Three objects could not be used for these
simulations; B14623, Bs013544 and Bs017524. In each case the light
from the nearby object prevent us from using the object image as a
template for the spectrum (see Figures \ref{Bstamps2} and
\ref{Bfillers}) when measuring the total aperture correction. In \S
\ref{cal}, we find that the slit losses are 0.15 mag. In our
simulations, we find the mean value to be $1.25 \pm 0.10$, larger than
our estimate of 1.15 based on one star.  The scatter of 0.10, or 10\%,
is close to, but smaller than, the estimate of 15\% from
\citet{stark2013}. We do not include this additional error in Table
\ref{spectab} but have added this to the fluxes in the figures.

These simulations assume that the K band light traces the star-forming
regions that generate the emission lines. The actual morphology of the
star-forming regions that dominate the line flux cause an additional
uncertainty in this aperture correction. \citet{Nelson2015} find, for
example, that for $z\sim1$ galaxies with stellar masses above
$10^{9.5}\ M_{\sun}$ that the length scale of the line emitting region
is 10\% larger than the stellar continuum, with a slight mass
dependence. The difference in the morphologies can lead to an
uncertainty in the aperture correction as large as 50\%
\citep{erb2006c,Yoshikawa2010}. However, note that these aperture
corrections do not affect the main result of this paper, which
concerns the ratios of emission lines.

\subsubsection{Additional Simulations}
\label{etcsim}

MOSFIRE has a sophisticated exposure time calculator. This computes a
spectrum specified by the user and places it on top of a real night
sky spectrum for Mauna Kea. We perform a set of calculations using
this software as a second, independent check on our error
estimates. We use the tool to generate artificial spectral lines that
matched our selection criteria and typical fluxes. We then place these
at random redshifts that match those of the range spanned by our
sample. At each redshift, we generated both [\ion{O}{3}] lines and
H$\beta$.  We assume a 200 minute exposure in 0\farcs 7 FWHM
seeing. We did this for two sets of fluxes, using typical flux ratios
as found in our sample. The background varies significantly across our
redshift range.  We mimicked our preselection (see \S \ref{presamp})
before plotting our simulations, which decreases the range of
signal-to-noise values possible. 

\begin{figure}
\begin{center}\includegraphics[width=.8\linewidth]{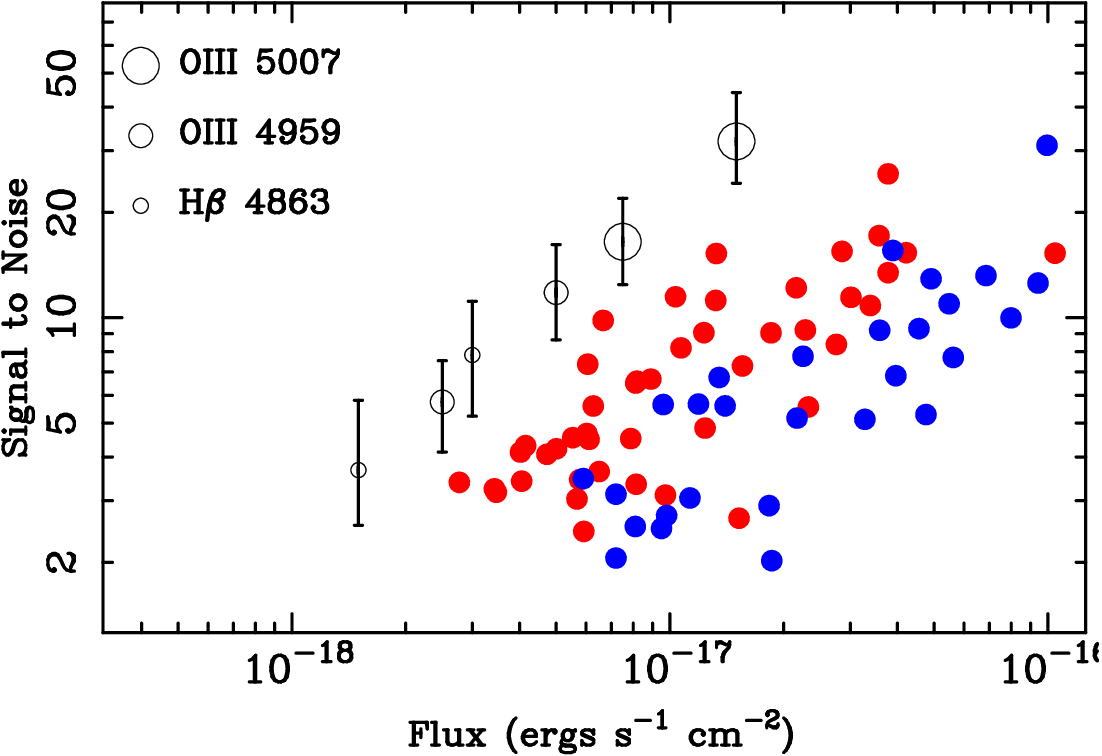}\end{center}
\caption{The signal to noise as a function of flux for our data (red)
  and the data from \citet{schenker2013} (blue). Also plotted (with
  open circles) are two sets of simulations using the MOSFIRE exposure
  time calculator for objects with typical fluxes spanning the
  redshift range of our sample. For these simulations, objects were
  assigned random redshifts in the range of our sample. At each
  redshift, we calculated the signal to noise using the exposure time
  calculator for a fixed flux for all three lines. Because the
  background varies, the resulting signal to noise changes. We plot
  the median values and show with error bars the range that
  encompasses 68\% of our simulations.  The simulated fluxes are total
  fluxes, while we plot the observed fluxes from our sample which do
  not include an aperture correction. Therefore, our data lie below
  the values from the exposure time calculator. \citet{schenker2013}
  find a lower signal to noise at a given flux, likely because of
  different observing conditions.}
\label{etc_err}

\end{figure}

We plot the results in Figure \ref{etc_err}. We plot the median values
with points and show the range covered by 68\% of the simulations with
error bars. The points with error bars are from our simulation and
represent an upper limit of what should be possible with the MOSFIRE
spectrometer, as there are no errors associated with extraction or
systematics in the sky subtraction. In general our data have S/N at a
given flux below the optimal results. For comparison, we plot the
results of \citet{schenker2013}, which are generally at a lower
signal-to-noise for the same flux level as our data. Likely, this is
because of different conditions during observing. We conclude that our
error estimates lie between the theoretical optimum and estimates from
other work.

\section{Analysis}
\label{sec:analysis}

We detected emission lines for all our primary target galaxies. We
also detected lines for three of our fainter LBGs that were secondary
targets. We note here we used no pre-selection for these secondary
targets.  The resulting line fluxes and redshifts are listed in Table
\ref{spectab}. For almost every spectrum, it was straightforward to
identify the [\ion{O}{3}] and H$\beta$ lines. For a few targets we do
not detect H$\beta$, usually because of night sky emission (see
Figures \ref{Bstamps}, \ref{Bstamps2}, and \ref{Bfillers}).

For two galaxies, Bs006541 and Bs009818, we additionally detect
H$\gamma$ in emission. The part of the night sky that this line falls
in, however, is full of Telluric absorption features and our flux
calibration is too uncertain in that region, which is why we do not
list flux measurements for these lines in Table \ref{spectab}.

The rest-frame optical redshifts tabulated in Table \ref{spectab} are
average values, weighted by the significance of the detected lines. We
find excellent agreement between our rest-frame optical redshifts and
the pre-existing rest-frame UV redshifts. The average redshift
difference is $\delta z = -0.0023 \pm 0.0009$. For the 8 Ly$\alpha$
emitters (LAEs) among our sample, this corresponds to a velocity
offset between Ly$\alpha$ and the [\ion{O}{3}] line of 153$\pm 60$ km
s$^{-1}$, possibly indicating somewhat lower outflow velocities than
in LAEs at $z\sim2$ \citep[see also][]{schenker2013}.

\subsection{Estimating the Equivalent Widths}

To measure the equivalent widths, we need an estimate of the continuum
in the \K-band for all our target galaxies. As none of our target
galaxies have reliably detected continua, we used the ground based \K\
image to estimate their continuum fluxes. The calibration for flux
loss we estimated in \S \ref{cal} is, in effect, a total flux in the
spatial direction but an aperture of 0\farcs 7 in the dispersion
direction because the data and calibrations are all measured in a
slit. We elected to measure the continua expected in the \K-band by
measuring the magnitude of each target galaxy and the star in a
circular aperture with a diameter of 0\farcs 7. This aperture
magnitude for the star is $K = 18.47$ mag AB. Thus, the difference
between the flux in the slit and the flux in the aperture is 0.23
mag. We add this offset to each of our \K\ measurements when
estimating the continuum to calculate the equivalent width.

Given the strengths of the rest-frame optical emission lines in the
\K-band, it is clear that these lines will contribute a significant
fraction to the total \K-band flux. In order to correct for this and to
estimate the clean \K-band continuum fluxes, we use a line-free
star bursting galaxy template from \citet{kinney1996}, to which we add
emission lines with the individual strengths that we measured (as
listed in Table \ref{spectab}). This template was then rescaled to
match the observed, aperture-corrected $K$ magnitude, listed in Table
\ref{sedtab}, for each target galaxy, from which we obtain the
normalization of the line-free continuum.

\begin{figure}
\begin{center}\includegraphics[width=0.8\linewidth]{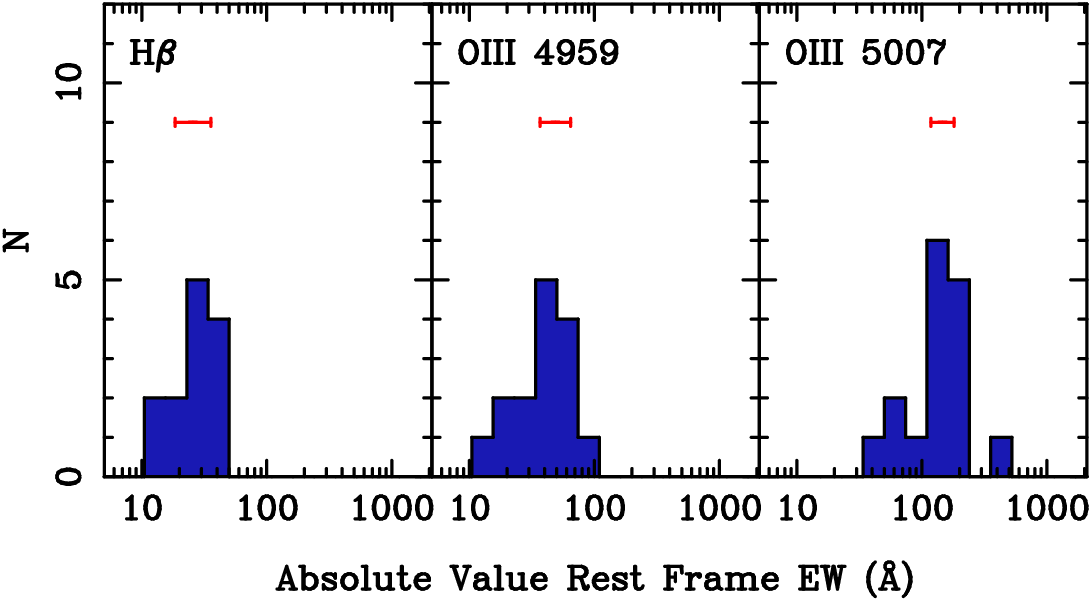}\end{center}
\caption{Distribution of the absolute value of the rest-frame
  equivalent widths for detected emission lines in our sample. Each
  panel separately shows the distribution of equivalent widths based
  on lines detected at $>\fac\sigma$, for H$\beta$ (left),
  [\ion{O}{3}] 4959 (middle), and [\ion{O}{3}]$\lambda$5007
  (right). We show with red error bars the median error and median
  equivalent width for each measurement. The error includes both a
  systematic term of 10\% and the error on the flux. The latter is generally a
  larger fraction of the EW for fainter objects, so the errors
  increase to smaller EWs. The systematic uncertainty of 10\% could be a
  lower bound, as other work has found values as large as 50\%, see
  Sec. \ref{apersim}}
\label{ews}
\end{figure}

In Figure \ref{ews}, we plot the rest-frame equivalent width
distribution including all lines detected at $>\fac\sigma$. As can be
seen, the H$\beta$ equivalent widths lie in the range
EW$_0($H$\beta) = 10\mbox{--}50$ \AA, while the
[\ion{O}{3}]$\lambda$5007 EW$_0$s are very strong with a median of
EW$_0 = 200$ \AA\ (i.e. $\sim$ 900 \AA\ observed-frame).

Our observations therefore confirm that high equivalent width
rest-frame optical lines result in a very large contribution of
emission lines to the rest-frame optical photometry of $z>4$ galaxies,
given the width of the IRAC bands of $\sim1~\mu$m. Ignoring this
effect can significantly bias the estimates of stellar masses in such
galaxies, and it is therefore very important to derive reliable
estimates of the equivalent width distribution of these rest-frame
optical lines for large samples of galaxies in the future \citep[see
e.g.][]{schenker2013}.

The errors in the equivalent widths are computed by first adding in
quadrature the errors in the aperture-corrected line flux and the
error in the $K$ continuum magnitude. We then add the systematic error
on the line flux from the aperture correction, 10\%, to the term. We
tabulate the equivalent width values along with the aperture-corrected
fluxes in Table \ref{ewtab}

\subsection{Properties from Broad Band Photometry: SFRs, Extinction, and Stellar Masses}
\label{broadprops}
Using the rest-frame ultra-violet data, we can estimate the UV
continuum spectral slope, $\beta$, which provides a measurement of the
extinction in star-forming galaxies via the IRX-$\beta$ relation
\citep[see][]{meurer1999}.  The UV slope $\beta$ is measured from a
power-law fit to the broad-band filters which sample the rest-frame
$1400\AA$ to $2800\AA$. This includes 4-5 filters from $V_{606}$ to
$J_{125}$, depending on the exact redshift of the sources. In Table
\ref{sedtab}, we list the measured UV spectral slope. With the dust
correction, we can then estimate a star-formation rate from our UV
imaging alone \citep[using the relations of][]{kennicutt1998}.  We
will call this dust-corrected SFR from the UV, SFR$_{\rm UV}$.

In order to estimate stellar masses for our galaxies, we use the
ZEBRA+ spectral energy distribution fitting code \citep{oesch2010}
with \citet{bc03} models at sub-solar metallicity ($0.2Z_\odot$) and
constant star-formation. We verified that the stellar masses do not
change significantly if using different assumptions for the
star-formation histories, such as exponentially increasing or
decreasing functional forms. By fitting models we can estimate
star-formation rates and the UV spectral slope ($\beta_{SED}$) which
are internally consistent with the stellar mass values. We find that
the values of $\beta_{SED}$ are systematically offset by 0.2 from the
$\beta$ we find from fitting a power law to the broad band photometry,
in line with differences seen in \citet{Finkelstein2012}. For the rest
of the paper, we will use the value of $\beta_{SED}$ unless we
specifically state otherwise, as the $\beta_{SED}$ estimate uses all of
the photometry, instead of a subset of four passbands. 

Given the large contribution of rest-frame optical emission lines to
the \K-band photometry, we self-consistently add both nebular emission
lines as well as nebular continuum emission to the templates. This is
done by converting 80\% of ionizing photons from these templates to
recombination lines for H and He using case B recombination
\citep{osterbrock2006} and adding metal lines relative to the H$\beta$
fluxes using the tabulated relations of \citet{anders2003}.

\section{Results and Discussion}
\label{sec:results}

\vspace{.3cm}
\subsection{Different Indicators of the Star-Formation Rate}

\begin{figure}
\begin{center}\includegraphics[width=.8\linewidth]{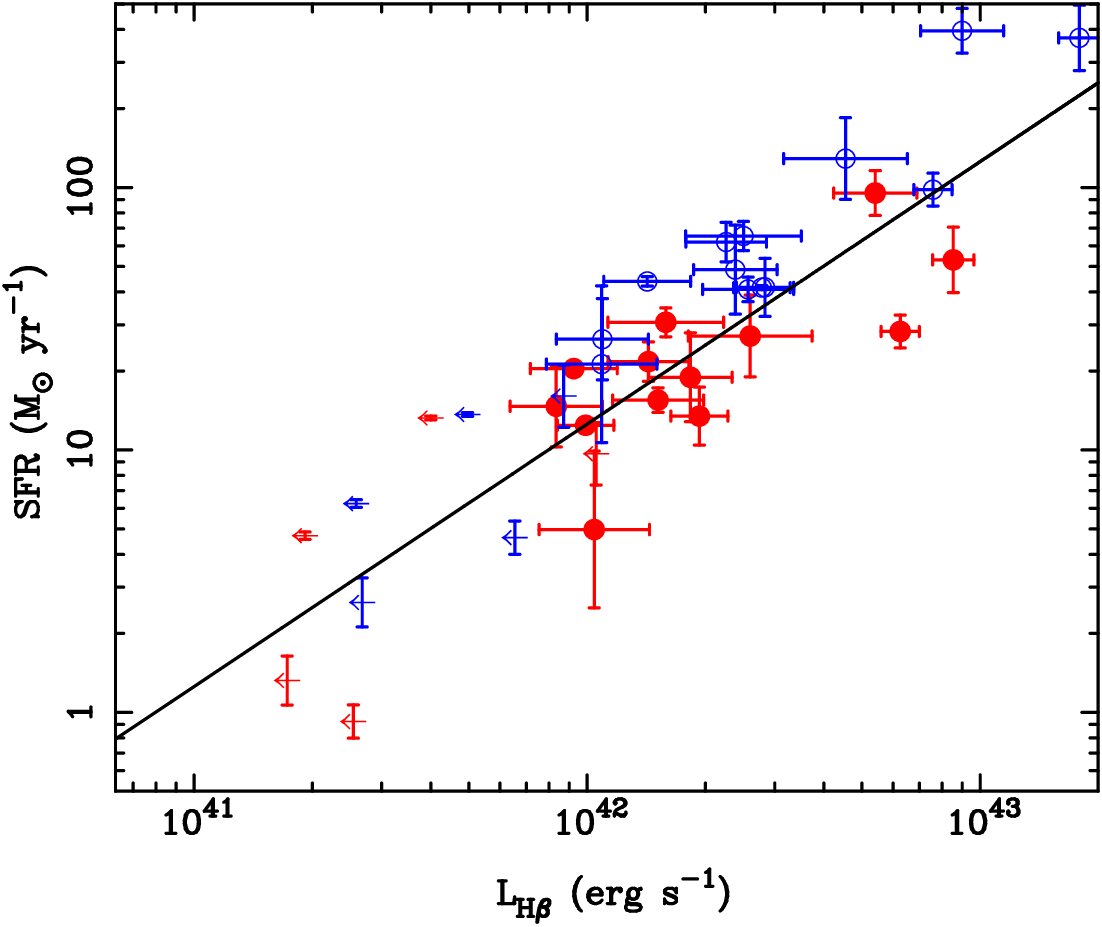}\end{center}
\caption{Distribution of the the H$\beta$ luminosity compared with our
  estimates of the dust-corrected UV star-formation rate (red) and
  star-formation rate derived from fitting spectral energy
  distributions (blue.) The luminosities are shown only for lines
  detected at $>\fac\sigma$. For comparison we plot the relation
  between the H$\beta$ luminosity and the star-formation rate from
  \citet{kennicutt1998}, after correcting the H$\beta$ luminosity from
  the H$\alpha$ luminosity assuming case B recombination. Both the
  star-formation rate and the H$\beta$ luminosity have been corrected
  for dust assuming the extinction relation from \citet{calzetti2000}
  with an $R_V = 4.03$ for the UV star-formation rates (red.) The
  H$\beta$ luminosities were corrected using the extinction from the
  best-fitting spectral energy distribution (blue) though also
  assuming the extinction relation from \citet{calzetti2000}. The
  observed H$\beta$ luminosities are in good agreement with the
  expectation from the dust corrected UV-based SFR, while the SED
  based SFR and extinction correction show an offset of 60\%. The
  uncertainty in the dust correction required could easily shift our
  H$\beta$ luminosities to higher values.  This would bring our values
  inline with our SED based star-formation rates. The large degree of
  uncertainty in the dust correction, assumptions about metallicities
  and the star-formation histories mean that either measurement could
  be in good agreement with the star-formation rates from H$\beta$
  luminosities.}
\label{sfrHb}
\end{figure}

We plot the luminosity in H$\beta$ and the UV star-formation rate in
Figure \ref{sfrHb}, along with simple relation between H$\beta$ and
star-formation rate from \citet{kennicutt1998}. This relation assumes
case B recombination to estimate the flux ratio of H$\beta$ to
H$\alpha$, a factor of 2.86, and converts the initial mass functions
to a Chabrier, another factor of 1.8 from the Salpeter assumed by
\citet{kennicutt1998}. Both the H$\beta$ luminosity and the UV derived
star-formation rate have been corrected for dust extinction assuming
the dust extinction from \citet{calzetti2000} with an $R_V = 4.03$.
The extinction was derived from A1600 (see \S \ref{broadprops})
assuming the Calzetti extinction law, with the stellar continuum
having only 0.44 of the extinction of the emission lines. The ratio of
the stellar continuum extinction to emission line extinction is highly
uncertain in high redshift galaxies, ranging from 0.4 to 1.0
\citep[e.g.,][]{Kashino2013,salmon2015}. This, in turn, causes a factor of
$\sim$2 uncertainty in our derived H$\beta$ luminosities
\citep[e.g.,][]{Yoshikawa2010}. In fact, \citet{debarros2015} finds
the extinction ratio has a star-formation rate dependence and that a
value of 0.44 is on the extreme end.

Our galaxies all have been fit with spectral energy distributions, and
from those fits we can derive both a value of $\beta$ and a
star-formation rate.  In marked contrast to the H$\beta$ and UV
star-formation rates, we find that the star-formation rates derived
from fitting spectral energy distributions from population synthesis
models are higher. Those models should be more self-consistent in
handling the extinction and dust from the galaxies. Even using the
extinction correction from the fits of the stellar populations, as we
did in Figure \ref{sfrHb}, our model based star-formation rates are
larger than the values derived from the H$\beta$ luminosities. This
disagreement was pointed out in \citet{castellano2014}. The SED
based star-formation rates are somewhat elevated on average, which is
likely a result of different assumptions about metallicities,
resulting in different UV spectral slopes for a given age stellar
population. In addition, we assume a fixed relation between the
extinction of the stellar populations and that of the H2 regions, but
this could be star-formation dependent, metallicity dependent,
incorrect, or all three.  For our results below, the most important
measurements are the stellar mass and the specific star-formation
rate. As such, we use the SED based estimates of the star-formation
rate as fitting one model to all of the photometry provides a
self-consistent estimate of the mass, star-formation rate and stellar
extinction.

\begin{figure*}
\begin{center}\includegraphics[width=.8\linewidth]{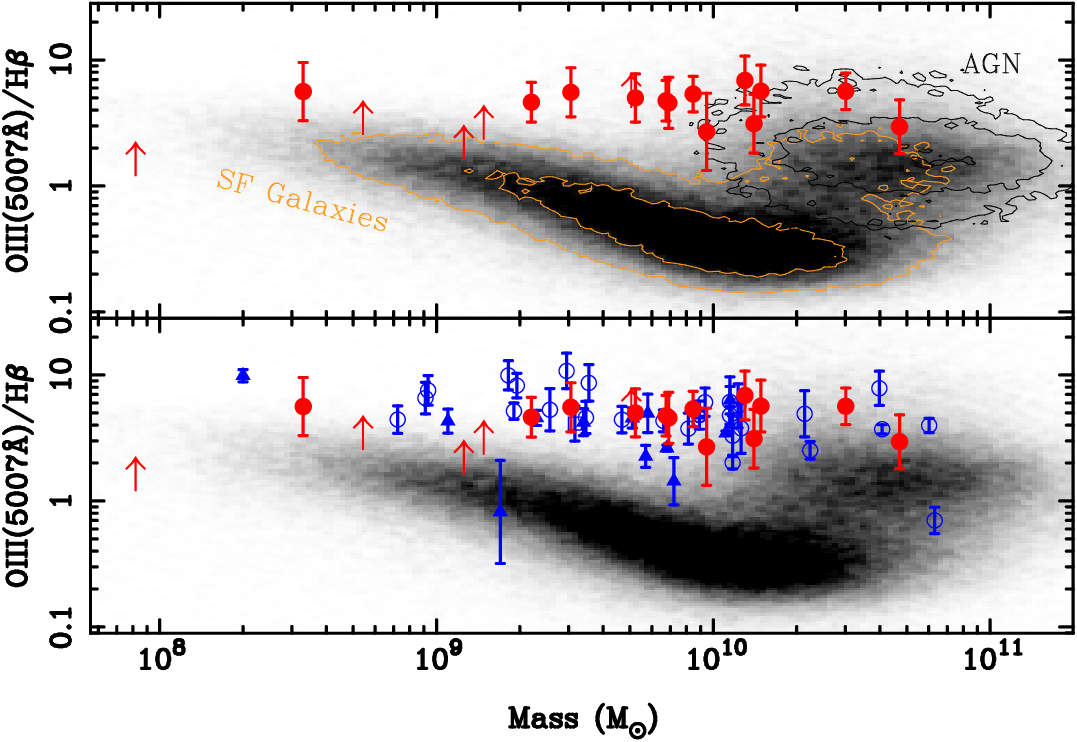}\end{center}

\caption{The ratio of the [\ion{O}{3}]$\lambda$5007 line to H$\beta$
  plotted as a function of stellar mass. Our data are shown in red,
  with upper limits denoted by arrows. In the lower panel, we also
  overplot the samples of \citet{troncoso2013} (blue open circles),
  and the sample of \citet{schenker2013} (blue solid triangles). We
  remove all detections of less than $\fac\sigma$ from
  \citet{schenker2013}. Evidently, all galaxies in our sample and from
  the literature show high ratios of [\ion{O}{3}] to H$\beta$, even at
  low stellar masses. For comparison, we show underlying grey scale
  from a sample of galaxies from the SDSS DR7 with line strengths as
  measured by \citet{brinchmann2004}. The stellar mass estimates are
  from the broad band photometry alone, mimicking the estimates of the
  higher redshift galaxies. In the upper panel, we overplot contours
  of the spectroscopic classifications from \citet{brinchmann2004},
  where the black contours show galaxies classified as AGN and the
  orange contours show galaxies classified as star-forming, based on
  the ratios of emission line strengths. Clearly, all high-redshift
  galaxies show line ratios significantly above the local star-forming
  galaxy population. Furthermore, the line ratios are essentially
  independent of stellar mass. }

\label{oxymass}
\end{figure*}

\subsection{The [\ion{O}{3}]/H$\beta$ Emission Line Ratios}

The primary goal of our observations was to analyze the emission line
properties of $z\sim3.5$ star-forming galaxies. In particular, there is
now growing evidence of increasingly high [\ion{O}{3}]/H$\beta$ line ratios
\citep[e.g.][]{kewley2013a} with redshift up to $z\sim3$, suggesting that the
conditions for star-formation might be quite different at high redshift
compared to local galaxies. With our sample of 18 galaxies with emission
line detections we can now further test these observations at
$z\sim3.5$.

In Figure \ref{oxymass} we plot the [\ion{O}{3}]/H$\beta$ ratio as a
function of mass. As can be seen, all our galaxies show line ratios
larger than 2, with a median of \frat, which appears to be independent
of stellar mass.

For a reference sample of typical local galaxies, we use the MPA-JHU
catalog of \citet{brinchmann2004}. This is not a directly comparable
sample, as the selection is very different, but provides a large number
of galaxies with stellar masses and emission line strengths. We compare
with a subset of the DR7 version of the \citet{brinchmann2004} catalog.
We restrict the redshift range to $0.015 < z < 0.08$. The resulting
catalog contains 260,647 galaxies covering a wide range of properties
and star-formation rates. Each galaxy has a measurement of the H$\beta$
and [\ion{O}{3}] flux and equivalent width. In addition, each galaxy has
an estimated stellar mass and star-formation rate. The stellar masses
come from the imaging data alone, and are thus comparable to our mass
estimates. 

The \citet{brinchmann2004} catalog also contains a classification for each
source, as either AGN driven or star-formation driven. We use this
classification to plot contours of AGN dominated sources and star-formation
dominated sources in Figure \ref{oxymass}. For our purposes, we
combine both AGN and star-forming classifications into one each. We
plot those galaxies classified as `composite' in Figure \ref{oxymass},
but they are ignored when generating the contours showing where AGN
or star-forming galaxies are distributed in the figure.

It is immediately clear from Figure \ref{oxymass} that the LBGs we have
observed lie far off of the relationship of star-forming galaxies in the local
universe. Many of the galaxies lie in the part of the diagram where sources are
classified as AGN in the local universe (illustrated by black contours). The
highest mass galaxies lie in a region dominated by AGN, while at lower masses
our sources lie in a region which is completely devoid of local galaxies.

In order to increase the sample size of high-redshift sources, we
additionally collect data from the literature. First, the work by
\citet{maiolino2008} and \citet{mannucci2010} provides measurements of
[\ion{O}{3}] and H$\beta$ line fluxes for a number of LBGs at similar
redshifts as our targets. We use the summary of data from
\citet{troncoso2013} which includes stellar masses and star-formation
rates as well as line strengths.  In fact, the galaxies CDFS-4414 and
CDFS-4417 from \citet{maiolino2008} and \citet{troncoso2013} are also
present in our sample as Bs012141 and Bs012208, respectively.

The sample of \citet{schenker2013} represents an
excellent combination with our data. These authors observed 20
galaxies with MOSFIRE that were selected essentially in the exact same
manner as our primary galaxy sample. Namely, they are LBGs, mostly
with pre-existing redshift measurements from rest-frame UV
spectra. Instead of the GOODS-S field, \citet{schenker2013} targeted
GOODS-N, however. For each of their galaxies \citet{schenker2013}
tabulate H$\beta$ and [\ion{O}{3}] flux measurements, along with
equivalent widths estimated from the $K$ band continuum. As they
tabulate the sum of the two [\ion{O}{3}] lines, we multiply their
tabulated line flux by 0.75 to estimate the strength of the 5007\AA\
line alone. \citet{schenker2013} include a 15\% calibration error for
their flux estimates. When we estimate the errors on the ratios of the
line fluxes, we remove that calibration error. 

All high-redshift measurements from the literature are shown as blue
symbols in the lower panel of Figure \ref{oxymass}, clearly showing
that a high ration of [\ion{O}{3}] over H$\beta$ is an ubiquitous
feature among star-forming galaxies at $z\sim3$. We find a mean ratio
of \fratall\ when we combine our sample with the rest from the
literature, after removing the two galaxies in common between our
sample and the sample of \citet{maiolino2008}. It is also clear from
the Figure, that even after the combination with the larger sample
from the literature, we find very little dependence of the line ratio
on the stellar mass.

\subsection{Possible Contribution by Active Galactic Nuclei}

The high observed line ratios in our $z\sim3.5$ galaxy sample trigger
the question, whether these are all dominated by active galactic
nuclei (AGN). In the local universe, we generally observe such high
values of the [\ion{O}{3}] to H$\beta$ ratio only in AGNs
\citep[e.g.][]{juneau2011}. \citet{trump2011} and \citet{trump2013}
find evidence for active galactic nuclei powering at least some of the
$z\sim1-2$ population and that the lower redshift mass-excitation
relations of \citet{juneau2011} still discriminate between AGN and
star-forming galaxies at $z\sim2$, after only small modifications to
slightly higher line ratios. Thus, the high [\ion{O}{3}] to H$\beta$
ratio we observe in Figure \ref{oxymass} could be, for the higher
stellar mass galaxies at least, a result of black hole accretion.
If the division remains the same at $z\sim3.5$ as at $z\sim1.5$,
almost all of the galaxies in Figure \ref{oxymass} would be classified
as possible AGN.

Given the extremely deep X-ray data over the GOODS-South field, we can
test for such AGN contamination in our sample. Using the Chandra 4~Ms
catalog of \citet{xue2011}, we search to see if any of our sources are
obvious X-ray emitters. Unsurprisingly, none of the X-ray sources are
within a point spread function half-width half-maximum of our
spectroscopic targets.

We then stack the images of \citet{xue2011}, to see if we detected an
average signal from the galaxies in our spectroscopic sample. When
performing this stacking, we removed any sources detected in the
\citet{xue2011} catalog. We did this by excising a 7 pixel by 7
pixel box centered on the source, where each pixel is 0\farcs 492 in
size. To build up a background sample, we also extract postage stamps
of random parts of the X-ray images. We required that our stamps have
no sources within 25 pixels of the stamp center. Thus, both our source
stacks and our background stacks will have similar levels of
contamination from undetected objects. As the Chandra point spread
function becomes larger with increasing radius from the pointing
center, we only included background regions with centers whose radii
were the same range as our sources.

For both the source and the background stack, we extract a 5 by 5
pixel region at the center from the 0.5 - 2.0 keV soft band image. In
our 200 background images, we found 955 counts for an average of 4.8
counts per image. We found 122 counts in our stack of 18 source images
for an average 6.8 counts per source. This yields a net of 2.0
counts. This is not, however, statistically significant. We would
expect, at random, a stack of 18 images to show 210 counts 17\% of the
time.

Because AGN luminosity is correlated with the stellar mass of the
galaxy, we additionally assembled a stack of the nine highest stellar
mass galaxies in our sample. This stack has an average of 8.7 counts per image, for
a net 3.9, which only happens 6\% of the time. This is still not
statistically significant.

We note also that larger samples of galaxies in this redshift range, selected
by both photometric and by spectroscopic redshifts, show statistically
significant emission, but at luminosities consistent with star-formation
\citep[e.g.][]{cowie2012}. This implies that for typical galaxies in this
redshift range, the dominant source of the X-ray emission is not an AGN, but
rather star-formation.

Furthermore, for the galaxies for which we do have pre-existing
rest-frame UV spectra, we confirmed that none of these sources show
any evidence for the presence of an AGN (e.g. based on [\ion{N}{5}],
[\ion{C}{4}], or broad Ly$\alpha$ emission).

\citet{juneau2014} argues that, because of the higher flux limits we
have in our LBG sample as compared with those from the SDSS, we are
biased in our selection such that we will be more likely to find only
high [\ion{O}{3}] to H$\beta$ ratio galaxies and likely all of them
are AGN. We miss lower [\ion{O}{3}] to H$\beta$ ratio AGN because of
selection effects, and also likely miss the other indications of the
AGN powering the large ratio of the emission lines. \citet{coil2015}
argue that, in the MOSDEF sample, the selection effects from
\citet{juneau2014} are not as important as indicated. Rather, the
shift to [\ion{O}{3}] to H$\beta$ observed in star forming galaxies
comes about because of evolution in those systems.

In summary, we do detect an excess of X-ray events, especially in the
higher mass galaxy stack, but it is not statistically significant.
Additionally, we do not find any evidence for AGN contamination based
on the rest-frame UV spectra. However, we cannot completely rule out
the idea that active galactic nuclei generate the high observed
emission line ratios at least for some of the galaxies in our sample,
in particular if they are optically thick for X-rays.

\subsection{The Conditions of Star-formation in LBGs
  at $z\sim3.5$.}

Star-formation can produce high [\ion{O}{3}] to H$\beta$ ratios as
seen in our sample. For example, the models of \citet{dopita2000} and
\citet{kewley2001} have star-forming regions in intense starbursts
producing [\ion{O}{3}] to H$\beta$ ratios of $\sim5$. These models do
not require especially low metallicity or high electron densities, but
they do require high ionization parameters. For example, the peak
[\ion{O}{3}] to H$\beta$ in the continuous star-formation model of
\citet{kewley2001} is 5.4 for a model with a metal abundance of
$Z=0.2 Z_\odot$ and an electron density of 350 $cm^{-3}$, while the
electron density of 10 $cm^{-3}$ model produces a ratio of 5.2. Both
models, however, require ionization parameters of
$3\times10^{8} {\rm cm\ s^{-1}}$, one to two orders of magnitude
higher than found in local star-forming galaxies. In fact, low
metallicity alone cannot explain the observed ratios, only higher
ionization parameters can. When including the impact of radiation
pressure in dense star-forming regions, the recent models of
\citet{yeh2013} and \citet{verdolini2013} can produce even higher
[\ion{O}{3}] to H$\beta$ ratios, beyond even what we observe for
integrated galaxy light.

In all these models, a high [\ion{O}{3}] to H$\beta$ ratio implies a
significantly lower \ion{O}{2} line strength, around 70 to 80\% of the
H$\beta$ strength. This is indeed observed in high redshift galaxies
\citep[see][]{nakajima2013}.

There are other physical mechanisms proposed to raise the ionization
parameter. \citet{stanway2014} suggests binary star populations can
produce significantly elevated [\ion{O}{3}] to H$\beta$ ratios for
stellar populations over $\sim$100 Myrs. \citet{steidel2014} shows
that a 50,000K blackbody produces the necessary spectrum, in good
agreement with the results of \citet{stanway2014}. Interestingly,
\citet{erb2014} finds that the evidence for a harder ionizing spectrum
also occurs in the Ly $\alpha$ emission. \citet{steidel2014},
\citet{masters2014}, \citet{shapley2015}, \citet{Jones2015},
\citet{sanders2015} and \citet{cowie2015} all find that the $z\sim2$ redshift
population requires a different abundance ratio of N/O. Interestingly
\citet{dopita2016} derives a new calibration of the N and O abundance
and finds that this N/O shift is not required. The combination of a
harder spectrum \citep[e.g.,][]{steidel2014}, or a higher electron
density \citep[e.g.,][]{sanders2015}, along with possibly enhanced
abundances can explain the location of high redshift galaxies in these
excitation diagrams without any AGN contribution.

The above models provide a physical explanation for the large emission
line ratios we observe, namely a combination of a large ionization
parameter, harder ionizing spectrum, and possibly radiation
pressure. This implies that the conditions of star-formation are very
different in these high redshift LBGs as compared with galaxies at the
same stellar mass in the local universe.

\begin{figure*}
\begin{center}
\includegraphics[width=.8\linewidth]{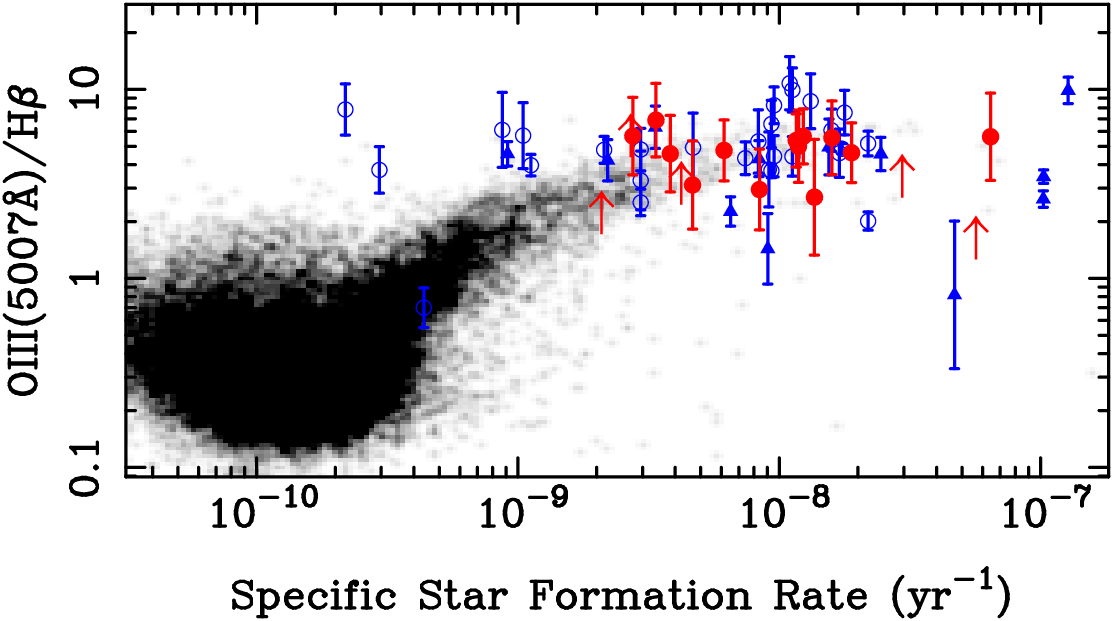}
\end{center}

\caption{The ratio of the [\ion{O}{3}]$\lambda$5007 line to H$\beta$
  plotted as a function of specific star-formation (the ratio of
  star-formation to stellar mass). Our data are shown in red, with
  upper limits denoted by arrows. The blue points represent data from
  \citet{schenker2013} (solid triangles) and \citet{troncoso2013}
  (open circles), excluding upper limits and the points with specific
  star-formations beyond the right edge of the plot.  For comparison,
  we plot a sample of galaxies classified as star-forming from the
  SDSS DR7 with line strengths as measured by \citet{brinchmann2004}
  as a grey scale. We remove all galaxies with star-formation rates
  below 2 \sfrun, the lowest star-formation rate of our high redshift
  sample. We note that we have purposefully chosen the limits of the
  grey scale to show the few galaxies in the local universe with the
  same properties as our high redshift sample. Clearly, all
  high-redshift galaxies show line ratios significantly above the
  typical local star-forming galaxy population.  However, a small tail
  of local galaxies extends to higher sSFR, which nicely connects up
  with the higher [\ion{O}{3}]$\lambda$5007 line to H$\beta$ ratios
  seen in the $z\sim3-4$ population
  \citep{brinchmann2008a,brinchmann2008b}.  Thus, the ionization
  and/or radiation pressure in local HII region appears to be driven
  by the overall specific star-formation, therefore coupling a local
  property with one on a galaxy-wide scale.}

\label{oxyssfr}
\end{figure*}

As was highlighted by \citet{brinchmann2008a} and \citet{brinchmann2008b},
galaxies with strong ionization parameters often have higher specific
star-formation rates than the typical star-forming galaxies in the local
universe. In Figure \ref{oxyssfr}, we plot the [\ion{O}{3}]/H$\beta$ specific
star-formation rate of galaxies in our sample, combined with the galaxies from
\citet{schenker2013} and \citet{troncoso2013}. We also show the location of
local galaxies with star-formation rates of at least 2 \sfrun\ and classified
as star-formation dominated by \citet{brinchmann2004}. While the typical
$z\sim0$ star-forming galaxy shows low ratios of [\ion{O}{3}]$\lambda$5007 to
H$\beta$ (only $\sim0.3$) and specific star-formation rates (SSFRs) of
$\sim10^{-10}$ yr$^{-1}$, a small tail of the local population extends to
significantly higher SSFRs and higher line ratios. Interestingly, this tail
nicely connects up with the location of the $z\sim3-4$ sample. This suggests
that (1) the ionization and/or the radiation pressure of HII regions is
connected with the global SSFR, coupling a local and a galaxy-wide property,
and (2) a small sub-sample of local star-forming galaxies exhibits similar
conditions of star-formation as the $z\sim3-4$ population. These are the
galaxies with similarly high SSFRs. Note, however, that these local galaxies on
average have a factor 4-5$\times$ lower masses than our $z\sim3-4$ sample.

\subsection{How representative is the current sample?}

As we have shown above, all of the observed LBGs have extreme star-forming
conditions, very different from typical local galaxies. Since LBGs are the
dominant contributors to the total star-formation rate at this high redshift
\citep{bouwens2009}, this suggests that star-formation was very different in
the early universe compared to the typical local galaxy. However, this depends
on how representative our sample is. The galaxies in our sample, and those in
\citet{troncoso2013} and \citet{schenker2013}, are generally selected by
Ly$\alpha$ emission.

\citet{shapley2003} found that the median Lyman $\alpha$ rest-frame
equivalent width for $z\simeq3$ galaxies was 0 \AA, implying that half
of the sample shows emission. Our sample is in agreement with that
fraction (8 out of 15 show Ly$\alpha$ emission). Combining our sample
with that of \citet{schenker2013}, however, we find 24 Ly$\alpha$
emitters out of 36 galaxies. By random chance, we would not expect to
draw 24 emitters out of 36 galaxies from the \citet{shapley2003}
sample. Thus it is likely that our the combined sample is biased
towards Ly$\alpha$ emitting LBGs.

When examining the properties of LBGs, \citet{shapley2003} also found
that those galaxies with Lyman $\alpha$ emission generally had lower
star-formation rates and lower amounts of dust extinction as measured
by bluer values of the UV continuum slope $\beta$. We find an average
star-formation rate for our sample of \sfrbar, well in-line with the
expectations of the Lyman $\alpha$ emitters in
\citet{shapley2003}. From this, we conclude that our sample, and the
ensemble of our sample with that of \citet{schenker2013}, resemble a
sample of moderate Lyman $\alpha$ emitters in properties.

In summary, we may not extend our results to the whole of the LBG 
population but, that being said, our sample combined with
that of \citet{schenker2013} seems typical for those galaxies with
modest Lyman $\alpha$ emission, characterized by lower star-formation
rates and lower dust content than typical LBGs.

\section{Conclusion}

We observed \ntot\ galaxies using the MOSFIRE spectrograph, where
\nspec\ were selected by having existing rest-frame UV redshift
measurements such that we could observe lines with minimal
interference from night-sky emission lines. The remaining nine
galaxies were $z\sim4$ LBGs that fell within the mask. For each
galaxy, we measured stellar masses, UV spectral slopes, and
star-formation rates using existing broad band imaging. From our
spectra, we measured the strength of the optical emission lines
H$\beta$ and the [\ion{O}{3}] lines at 4959 and 5007\AA.

Our main findings are:
\begin{itemize}

\item Every galaxy in our primary sample has a detected emission line.
  Three out of the {\bf nine} galaxies without a known redshift were
  also detected, pointing to the possibility of a more general survey of
  galaxies based on a photometric redshift selection alone.

\item The ratio of [\ion{O}{3}] to H$\beta$ is much higher, \frat\,
  than in similar mass star-forming galaxies in the local
  universe. When we combine our results with other $z\sim3$ samples,
  the ensemble of \ncomb\ galaxies has a median value of \fratall,
  unlike the values of 0.3 to 1 found in the local universe. This
  implies that the typical conditions for star-formation at $z\sim3.5$
  for UV bright galaxies are very different than in the local
  universe. These values require a combination of higher ionization
  parameters, higher electron density, a harder ionizing flux and a
  different N/O ratio to explain along with lower gas-phase
  metallicities for the HII regions.
  
\item The line ratio of [\ion{O}{3}] to H$\beta$ is strongly
  correlated with specific star-formation in the local universe. A tail
  of local galaxies with the highest sSFRs shows elevated line
  ratios similar to what we find for $z\sim4$ galaxies,
  thus linking our high-redshift sources with the physical conditions
  of $z\sim0$ galaxies. This correlation of sSFR and line ratio implies that
  local physics within star-forming regions appears to be correlated with the
  larger scale rate of star-formation across a broad range in galaxy mass
  scales.

\end{itemize}

\acknowledgments{ The authors would like to thank David Koo, Renske
  Smit, Mark Krumholz, and Max Pettini for useful discussions.  The
  authors wish to recognize and acknowledge the very significant
  cultural role and reverence that the summit of Mauna Kea has always
  had within the indigenous Hawaiian community.  We are most fortunate
  to have the opportunity to conduct observations from this mountain.
  Partial support for this work was provided by NASA through Hubble
  Fellowship grant HF-51278.01 awarded by the Space Telescope Science
  Institute, which is operated by the Association of Universities for
  Research in Astronomy, Inc., for NASA, under contract NAS 5-26555.
  Additionally, this work was partially supported by NASA grant
  NAG5-7697, and NASA grant HST-GO-11563.  This work was further
  supported in part by the National Science Foundation under Grant
  PHY-1066293 and the Aspen Center for Physics.  This research has
  made use of the NASA/IPAC Extragalactic Database (NED) which is
  operated by the Jet Propulsion Laboratory, California Institute of
  Technology, under contract with the National Aeronautics and Space
  Administration.  }

{\it Facility:} \facility{{\em HST} (ACS, WFC3)}, \facility{Keck I (MOSFIRE)},
\facility{VLT (ISAAC, HAWK-I)}, \facility{{\em Magellan}:Clay (PANIC)}

\begin{deluxetable}{lllccc|l}
\tablecaption{Measurements from the Spectra.}
\tablecolumns{7}
\tabletypesize{\tiny}
\tablehead{
\colhead{Id} &\colhead{$\alpha$} &\colhead{$\delta$} &
\colhead{$f_{H\beta}$} &
\colhead{$f_{4959}$} &
\colhead{$f_{5007}$} &
\colhead{z} \\ 
\colhead{} &\colhead{(J2000)} &\colhead{(J2000)} &
\colhead{($10^{-18}\ {\rm erg\ s^{-1}\ cm^{-2}}$)} &
\colhead{($10^{-18}\ {\rm erg\ s^{-1}\ cm^{-2}}$)} &
\colhead{($10^{-18}\ {\rm erg\ s^{-1}\ cm^{-2}}$)} &
\colhead{} \\ 
} 
\startdata
Bs006507 & 3:32:14.78 & -27:52:37.70 & 0.1 $\pm$ 1.8 &  2.4 $\pm$ 4.0 &  5.8 $\pm$ 1.7 &  3.39810 $\pm$ 0.04528  \\ 
Bs006516 & 3:32:14.79 & -27:50:46.50 & 3.9 $\pm$ 2.4 &  9.7 $\pm$ 3.1 &  27.6 $\pm$ 3.3 &  3.21583 $\pm$ 0.04928  \\ 
Bs006541 & 3:32:14.82 & -27:52:04.61 & 8.9 $\pm$ 1.3 &  13.2 $\pm$ 1.2 &  42.2 $\pm$ 2.8 &  3.47730 $\pm$ 0.02188  \\ 
Bs008202 & 3:32:17.43 & -27:52:01.22 & 0.8 $\pm$ 0.9 &  1.3 $\pm$ 0.9 &  5.9 $\pm$ 2.4 &  3.49227 $\pm$ 0.09104  \\ 
Bs008543 & 3:32:17.89 & -27:50:50.14 & 18.5 $\pm$ 2.0 &  37.8 $\pm$ 2.8 &  104.5 $\pm$ 6.8 &  3.47484 $\pm$ 0.01968  \\ 
Bs008802 & 3:32:18.28 & -27:51:58.91 & 5.7 $\pm$ 1.9 &  6.5 $\pm$ 1.8 &  15.2 $\pm$ 5.7 &  3.70487 $\pm$ 0.03714  \\ 
Bs009818 & 3:32:19.81 & -27:53:00.86 & 6.1 $\pm$ 1.4 &  10.3 $\pm$ 0.9 &  33.9 $\pm$ 3.1 &  3.70715 $\pm$ 0.03825  \\ 
Bs010545 & 3:32:20.97 & -27:50:22.35 & 4.0 $\pm$ 1.0 &  6.1 $\pm$ 0.8 &  22.8 $\pm$ 2.5 &  3.48041 $\pm$ 0.03325  \\ 
Bs012141 & 3:32:23.24 & -27:51:57.87 & 3.4 $\pm$ 1.1 &  3.5 $\pm$ 1.1 &  10.7 $\pm$ 1.3 &  3.47083 $\pm$ 0.05601  \\ 
Bs012208 & 3:32:23.34 & -27:51:56.87 & 7.9 $\pm$ 1.7 &  8.1 $\pm$ 2.4 &  23.2 $\pm$ 4.2 &  3.47271 $\pm$ 0.03522  \\ 
Bs013544 & 3:32:25.15 & -27:48:52.62 & 4.2 $\pm$ 1.0 &  8.1 $\pm$ 1.2 &  28.5 $\pm$ 1.8 &  3.48718 $\pm$ 0.03807  \\ 
Bs014828 & 3:32:26.76 & -27:52:25.91 & 6.0 $\pm$ 1.3 &  5.5 $\pm$ 1.2 &  30.1 $\pm$ 2.6 &  3.56387 $\pm$ 0.02811  \\ 
Bs016759 & 3:32:29.14 & -27:48:52.62 & 8.2 $\pm$ 1.2 &  13.3 $\pm$ 0.9 &  37.8 $\pm$ 1.5 &  3.60331 $\pm$ 0.03017  \\ 
Bs017378 & 3:32:29.93 & -27:49:28.28 & 4.7 $\pm$ 1.2 &  6.3 $\pm$ 1.1 &  21.6 $\pm$ 1.8 &  3.23547 $\pm$ 0.02700  \\ 
Bs017524 & 3:32:30.10 & -27:50:57.73 & 2.5 $\pm$ 1.8 &  4.1 $\pm$ 1.2 &  12.4 $\pm$ 2.6 &  3.69918 $\pm$ 0.06252  \\ 
 \noalign{\vskip .7ex} \hline \noalign{\vskip 1ex}
B14623 & 3:32:26.50 & -27:51:02.21 & 0.2 $\pm$ 0.4 &  0.5 $\pm$ 0.9 &  2.0 $\pm$ 1.7 &  3.18798 $\pm$ 0.06445  \\ 
B15573 & 3:32:27.64 & -27:50:59.68 & 6.7 $\pm$ 0.7 &  12.3 $\pm$ 1.4 &  35.8 $\pm$ 2.1 &  3.58419 $\pm$ 0.02541  \\ 
B17453 & 3:32:30.02 & -27:50:41.35 & 2.8 $\pm$ 0.8 &  5.0 $\pm$ 1.2 &  15.6 $\pm$ 2.1 &  3.56878 $\pm$ 0.04179  \\ 
\enddata
\label{spectab}
\end{deluxetable}

\begin{deluxetable}{lrrrllllll}
\tablecaption{Photometry and Photometric Derived Measurements}
\tabletypesize{\tiny}
\tablecolumns{10}
\tablehead{
\colhead{Id} &\colhead{$i_{775}$} &\colhead{K} &\colhead{Mass} &\colhead{$\beta$} &\colhead{$\beta_{\rm SED}$} &\colhead{SFR$_{\rm UV}$} &\colhead{sSFR$_{\rm UV}$} &\colhead{SFR$_{\rm SED}$} &\colhead{sSFR$_{\rm SED}$} \\ 
\colhead{} &\colhead{(mag AB)} &\colhead{(mag AB)} &\colhead{($10^{9}$\ \msun)} &\colhead{} &\colhead{} &\colhead{(\sfrun)} &\colhead{(Gyr$^{-1}$)} &\colhead{(\sfrun)} &\colhead{(Gyr$^{-1}$)} \\ 
} 
\startdata
Bs006507 & 26.93 $\pm$ 0.13 & 26.45$\pm$ 0.21 & 0.08 $\pm$ 0.08 & -2.55 $\pm$ 0.35 & -2.06 $\pm$  0.16 & 0.9 $\pm$ 0.1 & 11.4 $\pm$ 1.7 & 4.6 $\pm$ 0.7 & 57.2 $\pm$ 8.3   \\ 
Bs006516 & 23.84 $\pm$ 0.01 & 23.94$\pm$ 0.04 & 3.64 $\pm$ 0.30 & -2.41 $\pm$ 0.21 & -2.30 $\pm$  0.02 & 13.2 $\pm$ 0.2 & 3.6 $\pm$ 0.1 & 13.6 $\pm$ 0.2 & 3.7 $\pm$ 0.1   \\ 
Bs006541 & 23.95 $\pm$ 0.02 & 23.44$\pm$ 0.02 & 5.25 $\pm$ 0.46 & -2.35 $\pm$ 0.13 & -1.82 $\pm$  0.03 & 12.4 $\pm$ 0.2 & 2.4 $\pm$ 0.0 & 41.5 $\pm$ 0.8 & 7.9 $\pm$ 0.1   \\ 
Bs008202 & 25.11 $\pm$ 0.03 & 24.96$\pm$ 0.11 & 1.11 $\pm$ 0.27 & -2.40 $\pm$ 0.17 & -2.24 $\pm$  0.05 & 4.7 $\pm$ 0.2 & 4.2 $\pm$ 0.1 & 6.3 $\pm$ 0.2 & 5.6 $\pm$ 0.2   \\ 
Bs008543 & 23.62 $\pm$ 0.01 & 22.73$\pm$ 0.01 & 23.95 $\pm$ 3.17 & -1.62 $\pm$ 0.19 & -1.23 $\pm$  0.07 & 53.0 $\pm$ 15.2 & 2.2 $\pm$ 0.6 & 371.6 $\pm$ 106.7 & 15.5 $\pm$ 4.5   \\ 
Bs008802 & 23.96 $\pm$ 0.02 & 23.80$\pm$ 0.04 & 7.56 $\pm$ 1.16 & -1.70 $\pm$ 0.22 & -1.41 $\pm$  0.06 & 27.2 $\pm$ 9.7 & 3.6 $\pm$ 1.3 & 128.9 $\pm$ 46.2 & 17.0 $\pm$ 6.1   \\ 
Bs009818 & 24.38 $\pm$ 0.02 & 24.18$\pm$ 0.06 & 2.65 $\pm$ 0.53 & -1.92 $\pm$ 0.23 & -1.78 $\pm$  0.03 & 18.9 $\pm$ 7.4 & 7.1 $\pm$ 2.8 & 48.6 $\pm$ 19.0 & 18.3 $\pm$ 7.2   \\ 
Bs010545 & 24.59 $\pm$ 0.03 & 24.01$\pm$ 0.04 & 9.98 $\pm$ 3.46 & -1.72 $\pm$ 0.11 & -1.45 $\pm$  0.07 & 15.5 $\pm$ 1.7 & 1.6 $\pm$ 0.2 & 40.9 $\pm$ 4.4 & 4.1 $\pm$ 0.4   \\ 
Bs012141 & 24.19 $\pm$ 0.03 & 24.13$\pm$ 0.04 & 10.90 $\pm$ 1.60 & -1.61 $\pm$ 0.12 & -1.38 $\pm$  0.04 & 30.7 $\pm$ 3.9 & 2.8 $\pm$ 0.4 & 65.3 $\pm$ 8.3 & 6.0 $\pm$ 0.8   \\ 
Bs012208 & 23.40 $\pm$ 0.01 & 22.63$\pm$ 0.01 & 36.58 $\pm$ 2.99 & -1.41 $\pm$ 0.15 & -1.14 $\pm$  0.04 & 95.2 $\pm$ 18.7 & 2.6 $\pm$ 0.5 & 395.4 $\pm$ 77.7 & 10.8 $\pm$ 2.1   \\ 
Bs013544 & 23.96 $\pm$ 0.01 & 23.47$\pm$ 0.02 & 9.56 $\pm$ 1.31 & -2.00 $\pm$ 0.10 & -1.77 $\pm$  0.04 & 20.4 $\pm$ 0.8 & 2.1 $\pm$ 0.1 & 43.9 $\pm$ 1.8 & 4.6 $\pm$ 0.2   \\ 
Bs014828 & 23.95 $\pm$ 0.02 & 23.58$\pm$ 0.03 & 4.42 $\pm$ 0.74 & -1.99 $\pm$ 0.14 & -1.76 $\pm$  0.04 & 21.7 $\pm$ 3.7 & 4.9 $\pm$ 0.8 & 62.0 $\pm$ 10.7 & 14.0 $\pm$ 2.4   \\ 
Bs016759 & 24.54 $\pm$ 0.02 & 23.85$\pm$ 0.04 & 1.94 $\pm$ 0.20 & -2.01 $\pm$ 0.17 & -1.81 $\pm$  0.03 & 13.5 $\pm$ 3.4 & 6.9 $\pm$ 1.8 & 41.6 $\pm$ 10.6 & 21.5 $\pm$ 5.5   \\ 
Bs017378 & 24.05 $\pm$ 0.02 & 23.87$\pm$ 0.03 & 5.13 $\pm$ 0.56 & -2.03 $\pm$ 0.23 & -1.88 $\pm$  0.05 & 14.7 $\pm$ 5.2 & 2.9 $\pm$ 1.0 & 26.4 $\pm$ 9.4 & 5.1 $\pm$ 1.8   \\ 
Bs017524 & 25.15 $\pm$ 0.05 & 25.00$\pm$ 0.08 & 0.53 $\pm$ 0.09 & -1.92 $\pm$ 0.18 & -2.02 $\pm$  0.09 & 9.7 $\pm$ 2.6 & 18.3 $\pm$ 5.0 & 16.0 $\pm$ 4.4 & 30.4 $\pm$ 8.3   \\ 
\noalign{\vskip .7ex} \hline \noalign{\vskip 1ex} 
B14623 & 26.81 $\pm$ 0.15 & 26.49$\pm$ 0.38 & 0.88 $\pm$ 0.53 & -1.96 $\pm$ 0.14 & -1.73 $\pm$  0.13 & 1.3 $\pm$ 0.3 & 1.5 $\pm$ 0.3 & 2.6 $\pm$ 0.6 & 3.0 $\pm$ 0.6   \\ 
B15573 & 25.09 $\pm$ 0.03 & 23.60$\pm$ 0.03 & 6.76 $\pm$ 1.09 & -1.28 $\pm$ 0.13 & -1.18 $\pm$  0.02 & 28.3 $\pm$ 4.1 & 4.2 $\pm$ 0.6 & 98.2 $\pm$ 14.1 & 14.5 $\pm$ 2.1   \\ 
B17453 & 26.08 $\pm$ 0.07 & 24.75$\pm$ 0.08 & 0.32 $\pm$ 0.03 & -1.76 $\pm$ 0.38 & -1.73 $\pm$  0.04 & 5.0 $\pm$ 3.4 & 15.4 $\pm$ 10.6 & 21.2 $\pm$ 14.6 & 65.9 $\pm$ 45.3   \\ 
\enddata
\label{sedtab}
\end{deluxetable}

\begin{deluxetable}{lrrrrrr}
\tablecaption{Equivalent Width Measurements}
\tabletypesize{\tiny}
\tablecolumns{7}
\tablehead{
\colhead{Id} &\colhead{$f_{c,H\beta}$} &\colhead{EW$_{H\beta}$} &\colhead{$f_{c,4959}$} &\colhead{EW$_{4959}$} &\colhead{$f_{c,5007}$} &\colhead{EW$_{5007}$} \\ 
\colhead{} &\colhead{($10^{-18}\ erg s^{-1} cm^{-2}$)} &\colhead{(\AA)} &\colhead{($10^{-18}\ erg s^{-1} cm^{-2}$)} &\colhead{(\AA)} &\colhead{($10^{-18}\ erg s^{-1} cm^{-2}$)} &\colhead{(\AA)} \\ 
} 
\startdata
Bs006507 & 0.1 $\pm$ 2.2 & -27.2 $\pm$ 566.7 & 2.9 $\pm$ 5.3 & -782.8 $\pm$ 1419.3 & 7.2 $\pm$ 2.8 & -1920.6 $\pm$ 880.2 \\ 
Bs006516 & 4.9 $\pm$ 3.5 & -101.1 $\pm$ 73.1 & 12.2 $\pm$ 5.1 & -254.0 $\pm$ 107.6 & 34.5 $\pm$ 7.6 & -721.6 $\pm$ 162.9 \\ 
Bs006541 & 11.1 $\pm$ 2.8 & -149.1 $\pm$ 37.4 & 16.5 $\pm$ 3.1 & -221.9 $\pm$ 42.5 & 52.8 $\pm$ 8.7 & -710.6 $\pm$ 119.6 \\ 
Bs008202 & 1.0 $\pm$ 1.3 & -49.2 $\pm$ 59.8 & 1.6 $\pm$ 1.3 & -77.9 $\pm$ 61.4 & 7.4 $\pm$ 3.8 & -354.7 $\pm$ 185.4 \\ 
Bs008543 & 23.2 $\pm$ 4.9 & -171.8 $\pm$ 36.3 & 47.2 $\pm$ 8.2 & -351.6 $\pm$ 61.6 & 131.0 $\pm$ 21.6 & -973.9 $\pm$ 161.8 \\ 
Bs008802 & 7.1 $\pm$ 3.1 & -104.3 $\pm$ 45.0 & 8.1 $\pm$ 3.1 & -119.6 $\pm$ 45.3 & 19.1 $\pm$ 9.0 & -281.8 $\pm$ 134.2 \\ 
Bs009818 & 7.6 $\pm$ 2.5 & -162.5 $\pm$ 53.8 & 12.9 $\pm$ 2.4 & -275.8 $\pm$ 56.8 & 42.4 $\pm$ 8.2 & -905.5 $\pm$ 190.2 \\ 
Bs010545 & 5.0 $\pm$ 1.7 & -110.1 $\pm$ 38.0 & 7.6 $\pm$ 1.8 & -166.3 $\pm$ 40.1 & 28.5 $\pm$ 5.9 & -629.2 $\pm$ 135.6 \\ 
Bs012141 & 4.3 $\pm$ 1.7 & -95.3 $\pm$ 39.1 & 4.3 $\pm$ 1.8 & -96.5 $\pm$ 40.4 & 13.4 $\pm$ 3.0 & -299.2 $\pm$ 68.3 \\ 
Bs012208 & 9.8 $\pm$ 3.2 & -51.2 $\pm$ 16.5 & 10.2 $\pm$ 4.1 & -53.0 $\pm$ 21.2 & 29.1 $\pm$ 8.1 & -153.2 $\pm$ 42.9 \\ 
Bs013544 & 5.2 $\pm$ 1.7 & -65.2 $\pm$ 21.7 & 10.1 $\pm$ 2.6 & -128.5 $\pm$ 32.8 & 35.7 $\pm$ 5.9 & -454.3 $\pm$ 76.1 \\ 
Bs014828 & 7.5 $\pm$ 2.4 & -105.7 $\pm$ 33.5 & 6.9 $\pm$ 2.2 & -96.9 $\pm$ 31.2 & 37.7 $\pm$ 7.1 & -532.1 $\pm$ 102.4 \\ 
Bs016759 & 10.2 $\pm$ 2.6 & -189.5 $\pm$ 48.6 & 16.6 $\pm$ 2.8 & -308.0 $\pm$ 54.5 & 47.2 $\pm$ 6.5 & -879.0 $\pm$ 136.9 \\ 
Bs017378 & 5.9 $\pm$ 2.0 & -108.8 $\pm$ 37.8 & 7.8 $\pm$ 2.2 & -144.7 $\pm$ 40.7 & 27.0 $\pm$ 4.9 & -500.7 $\pm$ 93.9 \\ 
Bs017524 & 3.1 $\pm$ 2.5 & -139.0 $\pm$ 115.7 & 5.1 $\pm$ 2.0 & -230.5 $\pm$ 93.0 & 15.5 $\pm$ 4.8 & -706.6 $\pm$ 227.4 \\ 
\noalign{\vskip .7ex} \hline \noalign{\vskip 1ex} 
B14623 & 0.2 $\pm$ 0.6 & -46.4 $\pm$ 117.4 & 0.6 $\pm$ 1.1 & -122.1 $\pm$ 235.7 & 2.5 $\pm$ 2.4 & -504.0 $\pm$ 523.1 \\ 
B15573 & 8.3 $\pm$ 1.7 & -120.5 $\pm$ 24.9 & 15.4 $\pm$ 3.2 & -223.4 $\pm$ 47.8 & 44.7 $\pm$ 7.1 & -651.5 $\pm$ 107.9 \\ 
B17453 & 3.5 $\pm$ 1.4 & -158.6 $\pm$ 64.4 & 6.3 $\pm$ 2.1 & -287.4 $\pm$ 100.6 & 19.5 $\pm$ 4.6 & -894.4 $\pm$ 231.6 \\ 
\enddata
\label{ewtab}
\end{deluxetable}

\end{document}